\documentclass[12pt,a4paper]{article}
\usepackage{graphicx}
\def\be {\begin{equation}}
\def\ee {\end{equation}}
\def\beqy{\begin{eqnarray}}
\def\eeqy{\end{eqnarray}}

\def\as{\alpha_s}

\def\stop{\tilde{t}}
\def\stopbar{\overline{\tilde{t}}}
\def\stoponium{\stop \stopbar}

\begin{document}
\begin{titlepage}
\title{{\normalsize{\hfill PACS:13.60.Le;14.80.-j;14.80.Ly}}\\[25mm]
\bf Supersymmetric hadronic bound state  detection at $e^+e^-$ colliders}
\author{M. Antonelli\footnote{Mario.Antonelli@lnf.infn.it}
\\INFN Milano and \\Milano University , via Celoria 16, Milano, Italy \\
N. Fabiano\footnote{Nicola.Fabiano@lnf.infn.it} \\INFN National
 Laboratories, P.O.Box 13, I00044 Frascati, Italy}

\date{}
\maketitle
\begin{abstract}
We review the possibility of formation for a bound state made out of a stop
quark and its antiparticle. The detection of a signal from its decay
has been investigated for the case of a $e^+e^-$ collider.
\end{abstract}
\end{titlepage}

\section{Introduction}
In the Standard Model it has been verified that there is creation of bound
states for every quark but the top (see for 
instance~\cite{FADIN,PRR,KUHNTOP,ME,NOI} and references therein). The 
latter possibility is
ruled out due to the high value of the top quark mass, which is responsible for
its short lifetime.
The natural step forward would be to consider the possibility of bound states
creation outside the Standard Model. In this case we focus our attention to the
supersymmetric extensions of the Standard Model \cite{HABER},
in particular to the resonant production~\cite{DREES,MODRITSCH} and detection
of a bound state (supermeson) created from a stop and an anti--stop
(``stoponium'') at $e^+e^-$ colliders.


\section{Bound States}

In this Section we will review the bound states creation. For the SUSY case,
our assumption will be that
the bound state creation does not differ from the SM case, as the relevant
interaction is again driven by QCD, and is regulated by the mass of the
constituent (s)quarks.

A formation criterion states that \cite{NOI} the formation of a hadron can occur
only if the level splitting between the lying levels of the bound states, which
depend upon the strength of the strong force between the (s)quarks and their
relative distance \cite{ME}, is larger than the natural width of the state. It
means that, if
\be
\Delta E_{2S-1S} \ge \Gamma
\label{eq:critde}
\ee
where $\Delta E_{2S-1S}=E_{2S}-E_{1S}$ , $\Gamma$ is the width of the would--be
bound state, then the bound state exists. 

For the case of a scalar bound state $\stoponium$ , without referencing to a
particular supersymmetric model, we should consider the Coulombic two--body
interaction
\be
V(r)= -\frac{4}{3} \; \frac{ \alpha_s}{r}
\label{eq:coulomb}
\ee
with the two--loop expression for $\as$ \cite{2LOOPS}
\be
\alpha_s(Q^2)= \frac{4 \pi}{\beta_0 \log \left [ Q^2/\Lambda^2_{\overline{MS}}
\right ] } \left \{ 1-\frac{2 \beta_1}{\beta_0^2} \frac{\log \left [ \log
\left [ Q^2/\Lambda^2_{\overline{MS}} \right ] \right ]}{\log \left [ 
Q^2/\Lambda^2_{\overline{MS}} \right ] } \right \}
\label{eq:2loop}
\ee
with $ \beta_0=11-\frac{2}{3} \, n_f \;,\; \beta_1=51-\frac{19}{3} \, n_f$ .
Due to the present limits on the  stop mass \cite{PDG,ALEPH_STOP} 
we could either assume that the stop is lighter than the top quark,
that is $n_f=5$, or heavier, i.e. $n_f=6$.
The $\as$ expression (\ref{eq:2loop}) has to be evaluated at a fixed scale
$Q^2=1/r_B^2$ , where $r_B$ is the Bohr radius
\be
r_B= \frac{3}{4 \mu \alpha_s}
\label{eq:rbohr}
\ee
and $\mu$ is the reduced mass of the system.
It has been shown in \cite{NOI,ME} that in the case of  high quark mass values,
the predictions of the Coulombic potential evaluated at this scale
do not differ from the other potential model predictions.

\begin{figure}[p]
\begin{center}
\includegraphics[angle=-90,width=0.65\textwidth]{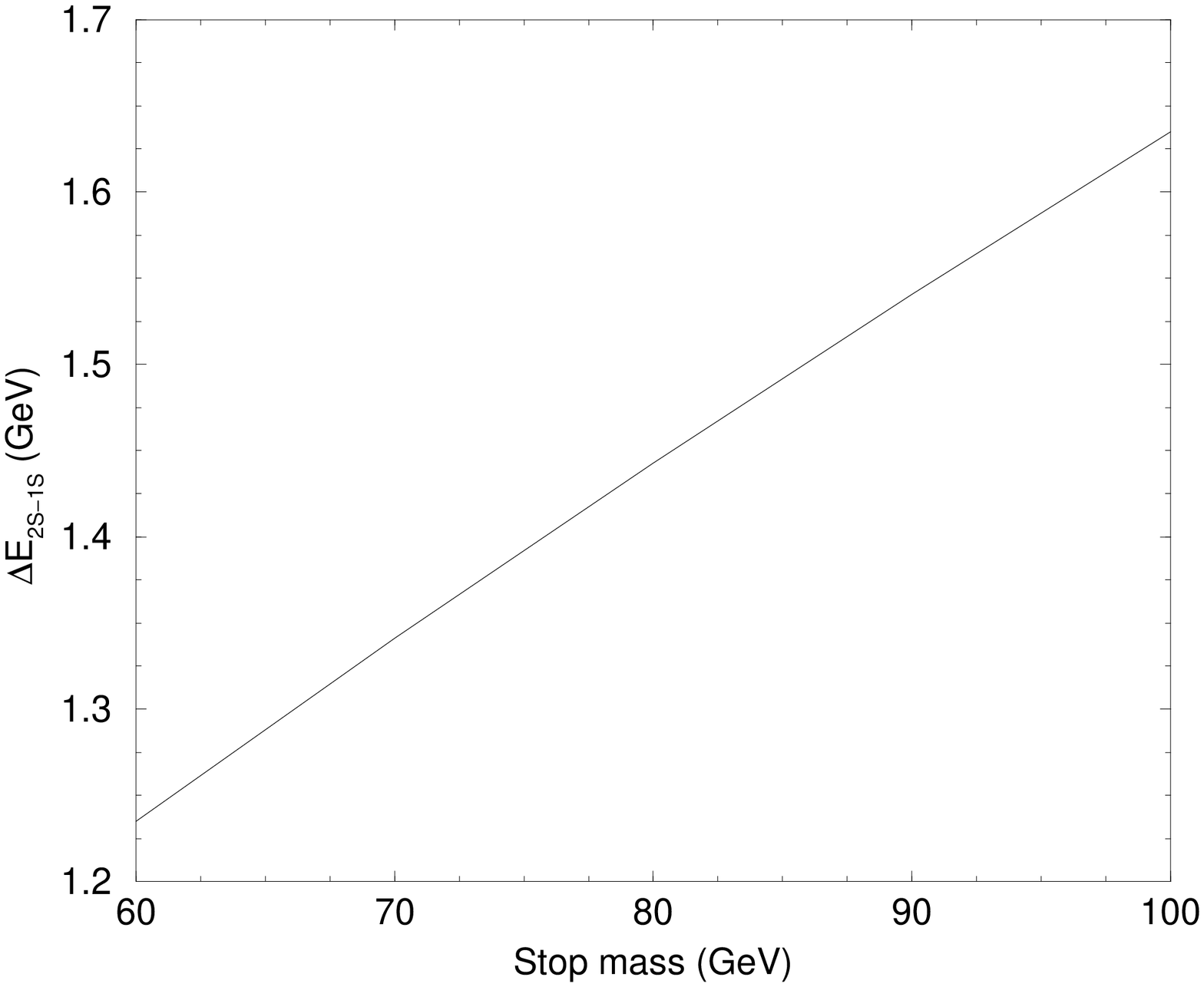}
\end{center}
\caption{ {\em $\Delta E_{2S-1S}$ as a function of the stop mass up to 100 GeV
for the Coulombic model.}}
\label{fig:stop1}
\end{figure}

\begin{figure}[hb]
\begin{center}
\includegraphics[angle=-90,width=0.65\textwidth]{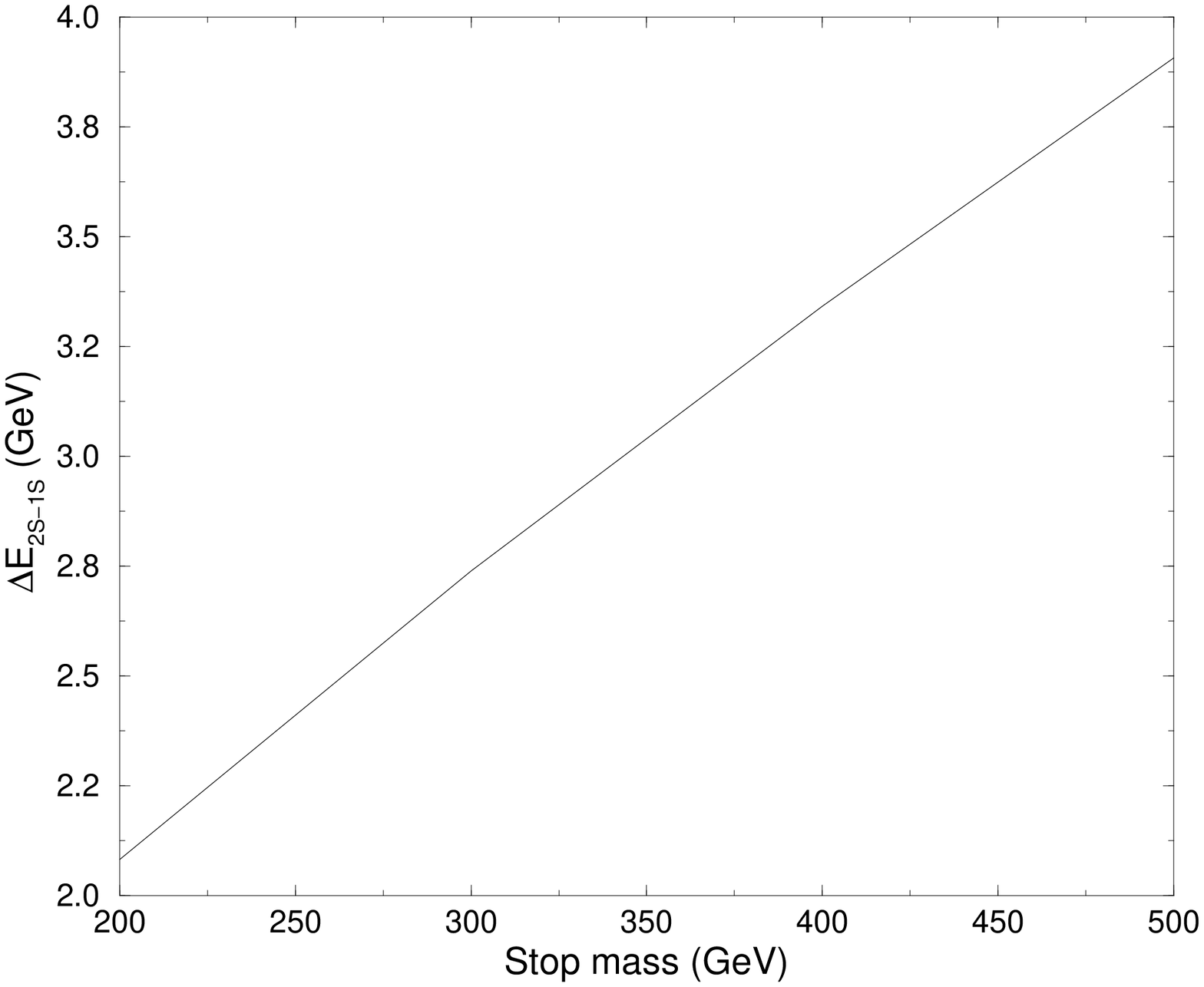}
\end{center}
\caption{ {\em $\Delta E_{2S-1S}$ as a function of the stop mass up to 500 GeV
for the Coulombic model.}}
\label{fig:stop1.nlc}
\end{figure}

In figures \ref{fig:stop1} and \ref{fig:stop1.nlc} we show a plot of the 
energy
splitting for the first two levels of the stoponium bound state with respect 
to the stop mass, for the LHC  and the NLC case respectively. As from
(\ref{eq:critde}), those figures have to be compared to the width of the 
stoponium.
The width of the stoponium, $\Gamma_{\stoponium}$ , is twice the width of the
single stop squark, as each should decay in a manner independent from the other.

There are several ways a stop should decay \cite{DECAY}, depending on the
assumptions made for the other superpartners. For very low values of the
stop quark mass, the highest width value will not exceed a few $KeV$, quite
smaller than the energy splitting of the first two levels of the stoponium. As
the mass increases more decay modes enter in and the width increases. In
particular  for the regime where $m_W + m_{\tilde{\chi}^0}
+ m_b < m_{\stop} < m_{\tilde{\chi}^+} +m_b$ the three body decay 
$\stop \to bW \tilde{\chi}^0$ is kinematically allowed and is comparable to the
flavour changing two body decay $\stop \to c \tilde{\chi}^0$ \cite{3BODY}.
Here $\tilde{\chi}^0$ refers to the lightest supersymmetric particle (LSP); 
$\tilde{\chi}^+$ is the lightest chargino. Even in this case those widths do 
not exceed values in the $KeV$ range. In this
scenario we see, as before, that the energy splitting is much larger than the
decay width of the bound state, thus hadronization is possible.

For even higher stop masses, the picture changes \cite{2BODY} as more
two body decays like $\stop \to b \chi^+$ and $ \stop \to t {\tilde \chi}^0$ 
are available. For these values of the stop mass there are regions of the
parameter space where the decay widths, even if lowered by the one--loop
corrections \cite{2BODY}, could overtake the energy levels splitting, thus
jeopardizing the formation of the supersymmetric bound state. For instance, in
the region where $\mu \sim M_2$ the decay width would be larger than $\Delta
E_{2S-1S}$ for stop masses of about 200 $GeV$, spoiling hadronization 
for $m_{\stop}$ beyond this range (here $\mu$ is the Higgs--higgsino mass 
parameter, while $M_2$ is the wino mass parameter).
On the contrary, for parameter values where $\mu \ll M_2$ , the decay width of
those modes are substantially lower. This would allow stoponium formation
for stop mass values in the energy range of the future NLC collider.
The region where $\mu \gg M_2$ is in a situation intermediate between the two
described above.

\begin{figure}[p]
\begin{center}
\includegraphics[angle=0,width=0.95\textwidth]{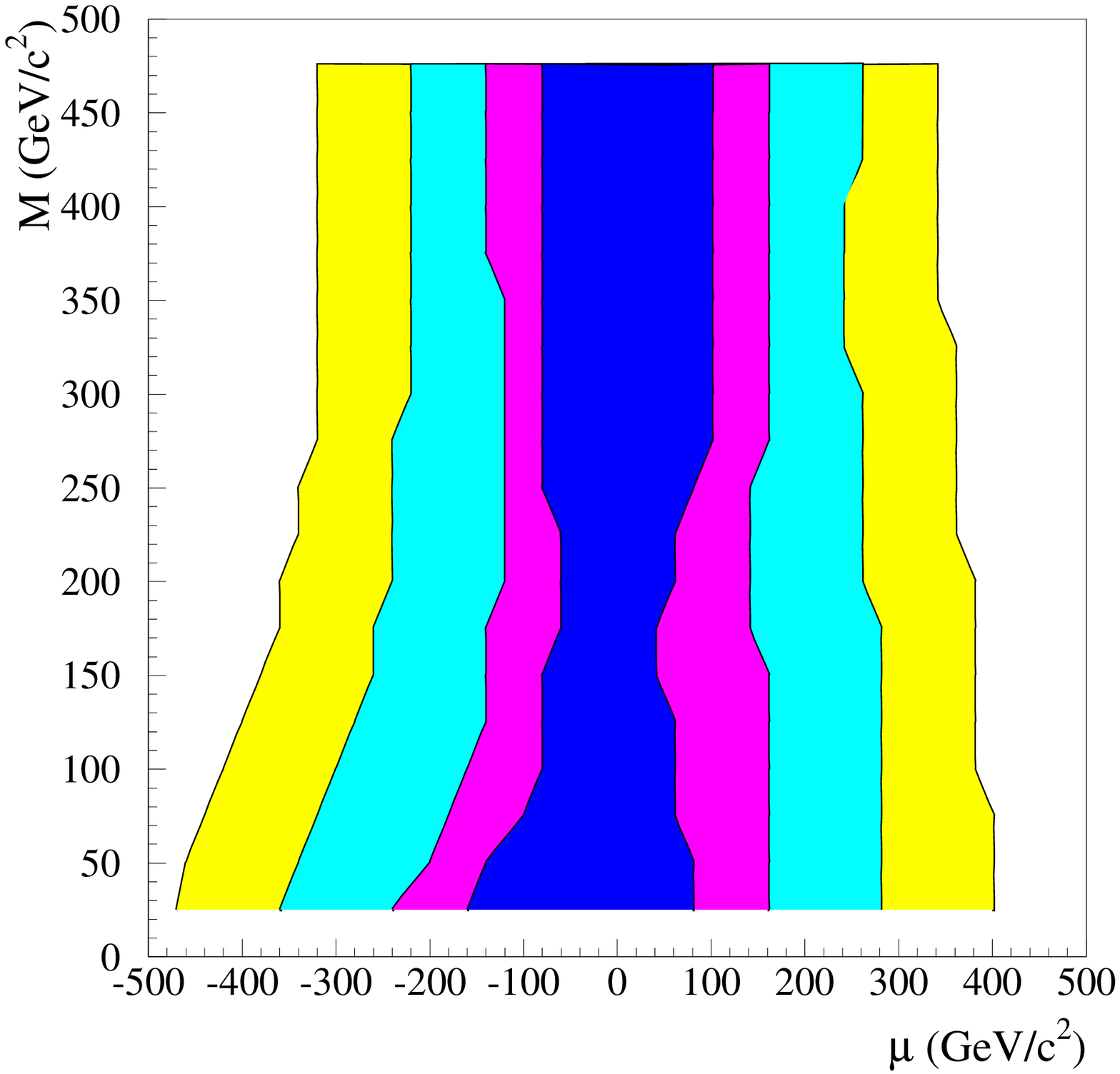}
\end{center}
\caption{ {\em Regions in the $\mu-M_2$ plane where stoponium  formation is
forbidden ;  $\tan \beta=1.5$.
The different colours refer to various values of the stop mass: 250, 300, 400
and 500 $GeV$ respectively, in increasing brightness.}}
\label{fig:range1}
\end{figure}

\begin{figure}[p]
\begin{center}
\includegraphics[angle=0,width=0.95\textwidth]{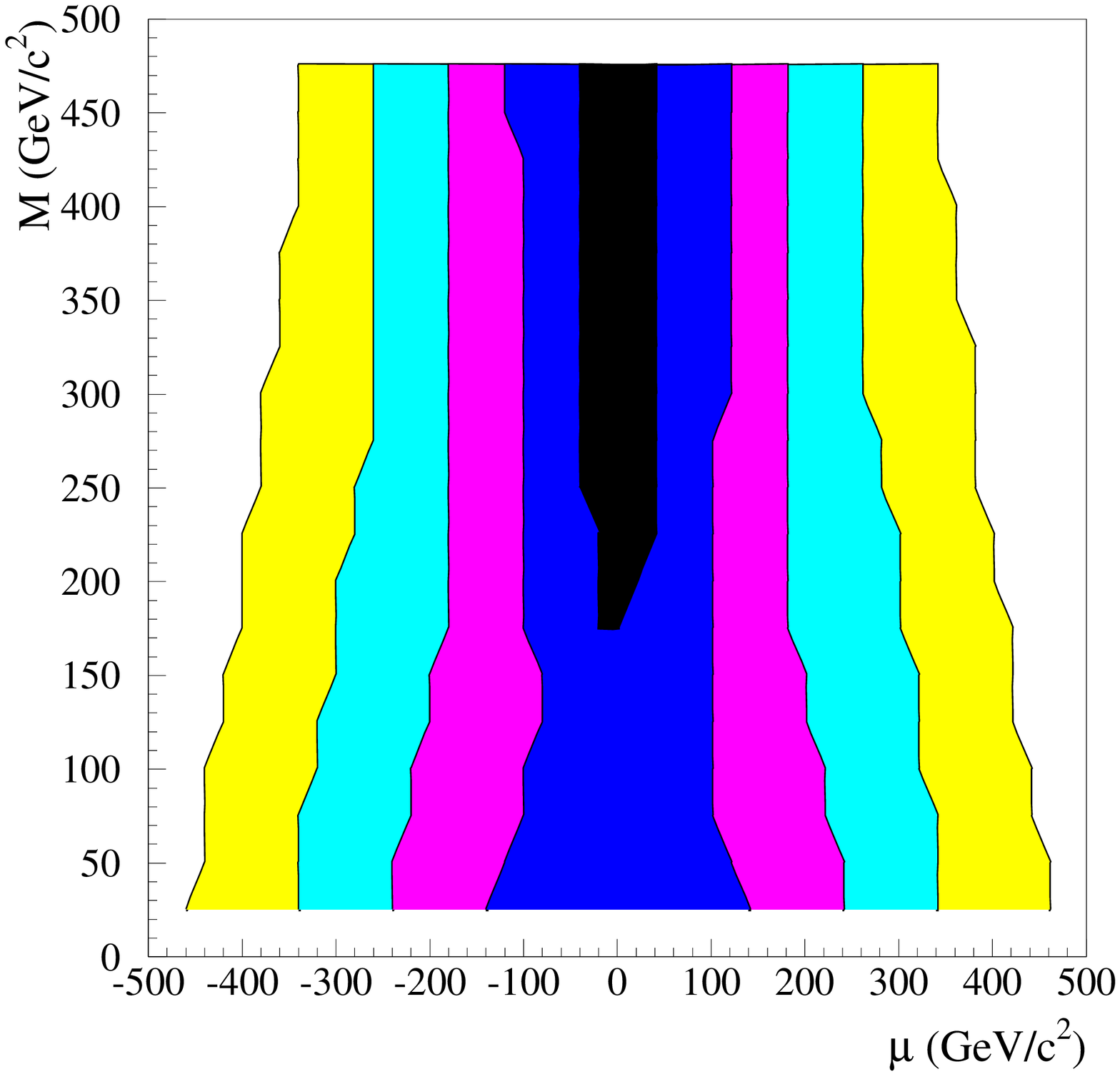}
\end{center}
\caption{ {\em  Regions in the $\mu-M_2$ plane where stoponium  formation is
forbidden ;  $\tan \beta=40$.
The different colours refer to various values of the stop mass: 100, 200, 300, 400
and 500 $GeV$ respectively, in increasing brightness.}}
\label{fig:range2}
\end{figure}

A quantitative description of the stoponium formation could be seen in 
figures~(\ref{fig:range1}) and~(\ref{fig:range2}), where we report the regions
of the $\mu-M_2$ plane for two values of $\tan \beta$ in which stoponium 
cannot be formed, as a function of the stop mass.

Regarding the hadronization problem we see that there are 
many possibilities due to the vast parameter space. For stop mass 
values under about 100--200 $GeV$ and $\tan \beta=1.5$
there is a window of opportunity for stoponium formation
regardless of the parameter values; beyond that range the stoponium 
formation would either be allowed or forbidden depending upon the
choice of the parameters.

\section{Cross Section and Decay Width}
The next natural step would be to see whether the stoponium could be 
detected on an $e^+e^-$
collider with LEP or future NLC characteristics. For this purpose we shall 
calculate its cross section and decay modes;
basing our predictions on \cite{NAPPI}, and updating their results.

We should look for the production and decay of the $P$ wave state, since we are
interested in the search of the bound state at a $e^+e^-$ collider, conserving
thus quantum numbers. 

We use the Breit--Wigner formula to evaluate the total cross section~\cite{PDG}:
\be
\sigma=\frac{3 \pi}{M^2} \times
\frac{\Gamma_e \Gamma_{tot}}{(E-M)^2+\Gamma_{tot}^2/4} 
\label{bw}
\ee
 where $M$ is the mass of the resonance, $E$ is the centre--of--mass energy, 
$\Gamma_{tot}$ is the total width, and $\Gamma_e$ is the decay width to
electrons.

The first decay we will investigate is the leptonic one,
which is given by the Van Royen--Weisskopf formula \cite{VANROYEN}
\be
\Gamma(2P \to e^+e^-) = 24 \alpha^2 Q^2 \frac{|R'(0)|^2}{M^4}
\label{eq:vanroyen}
\ee
$R'(0)$ is the derivative of the radial wavefunction calculated at the origin,
$M$ the mass of the bound state, $\alpha$ the QED constant, $Q$ the (s)quark
charge. In this case we are neglecting the stop coupling to the $Z$ boson, and
this allows to hide into the total width all the dependencies of the MSSM 
parameters for the cross--section formula~(\ref{bw}).

For this and following cases, we shall make use of the radial wavefunctions 
of the Coulombic model, as presented in Section (1). Those are, for the 
$1S$ state
\be
R_{1S}(r) = \left ( \frac{2}{r_B} \right )^{3/2} 
\exp \left (-\frac{r}{r_B} \right )
\label{eq:coul1s}
\ee
and for the $2P$
\be
R_{2P}(r) = \frac{1}{\sqrt{3}} \left ( \frac{1}{2r_B} \right )^{3/2} 
\frac{r}{r_B} \; \exp \left (-\frac{r}{2r_B} \right )
\label{eq:coul2p}
\ee
$r_B$ is the Bohr radius defined in (\ref{eq:rbohr}) .

For the hadronic width decay we have the following expression
\be
\Gamma(2P \to 3g) = \frac{64}{9} \; \as^3 \frac{|R'(0)|^2}{M^4} 
\log(m_{\stop} r_B)
\label{eq:ghad}
\ee
where the Bohr radius acts as an infrared cutoff \cite{NAPPI}.

\begin{figure}[p]
\begin{center}
\includegraphics[angle=-90,width=0.65\textwidth]{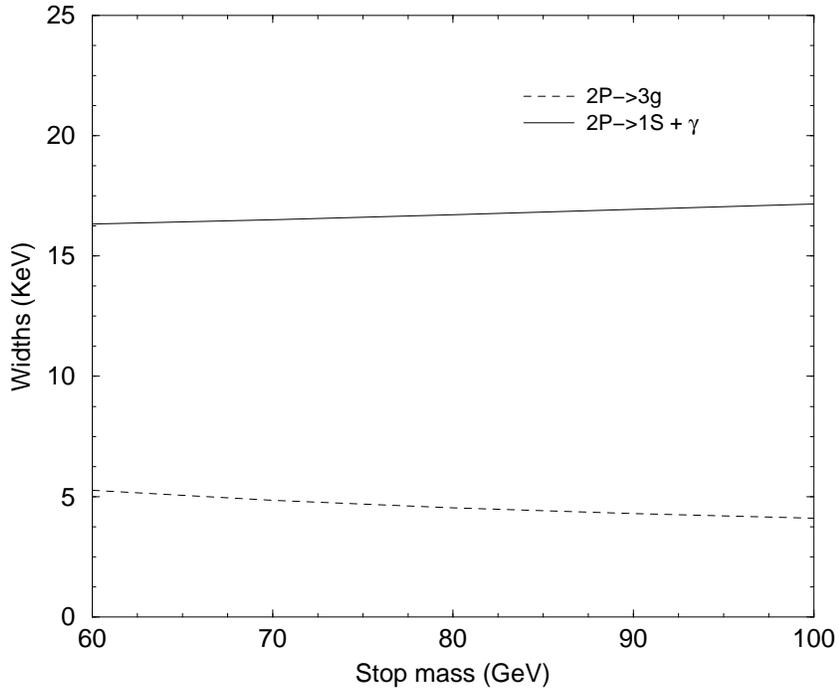}
\end{center}
\caption{ {\em Decay widths for the 2P state with respect to the stop mass 
for the Coulombic model. The dashed line represents the decay into hadrons,
the continuos line the decay into the $1S$ state and an emitted photon.}}
\label{fig:stop2}
\end{figure}

\begin{figure}[hb]
\begin{center}
\includegraphics[angle=-90,width=0.65\textwidth]{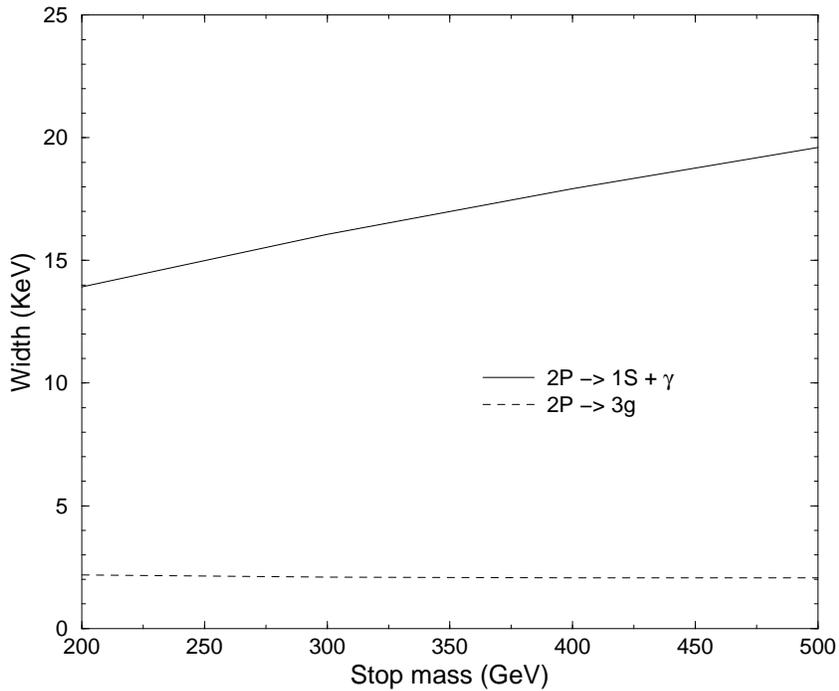}
\end{center}
\caption{ {\em Like Fig.~\ref{fig:stop2}, for a mass range of up to 
500 GeV, for NLC.}}
\label{fig:stop2.nlc}
\end{figure}

The $2P$ state could also decay into a $1S$ state and emit a photon. The 
width decay in this case is given by
\be
\Gamma(2P \to 1S + \gamma) = \frac{4}{9} \; \alpha Q^2 (\Delta E_{2S-1S})^3
D_{2,1}
\label{eq:g12gam}
\ee
where $\Delta E_{2S-1S}$ is the energy of the emitted photon, and
$D_{2,1}= \langle 2P | r | 1S \rangle$ is the dipole moment~\cite{KUHN}.
In figures \ref{fig:stop2} and \ref{fig:stop2.nlc} we present the decays of 
the $2P$ state into 
hadrons and into a $1S$ state plus a photon as a function of the stop mass,
as predicted by the Coulombic model. In this case the behaviour of the hadronic
decay width with respect to the stop mass is $\Gamma(2P \to 3g) \sim m\as^8 $~,
while the radiative decay width goes like $\Gamma(2P \to 1S + \gamma)\sim
m^2\as^5$~. In the former case the linear growth with $m$ is suppressed by the
high power of $\as$~, resulting in an essantially constant width for the stop
mass range of our interest. The $3g$ width will eventually grow faster for stop
mass values larger than about 1 $TeV$. The behaviour of the $2P\to 1S + \gamma$
decay is more straightforward, since it grows faster with $m$ and contains a
lower power of $\as$~. It is also apparent that among the two the $2P \to 1S +
\gamma$ decay width  dominates for increasing values of the stop mass as
it is clearly seen in figure~\ref{fig:stop2} and particularly
in~\ref{fig:stop2.nlc}.
It is possible to notice also a small
threshold effect due to the inclusion of the top flavour.

We must point out that this behaviour of decay widths of the stop bound
state is given by the particular Coulombic potential model used in the
computation. The results obtained however do not lose validity because, as it
has been shown in ~\cite{PRR,ME,NOI}, this Coulombic model does not differ 
significantly from other more popular potential models when the mass of the
constituent (s)quarks gets larger. 
This fact could be intuitively understood by considering the Bohr radius of the
bound state, which decreases like $1/m$: therefore the constituent (s)quarks
``feel'' more and more the Coulombic part of the potential which becomes
dominant with respect to other components of the potential, like for instance
the linear confining term that is added in the description of mesons containing
lighter quarks.

For a light stop the analysed annihilation modes are the dominant 
widths~\cite{DECAY} so far. As the stop mass increases the single stop decay
modes will dominate the total width because of the opening of the decay channel
$\stop \to t \tilde{\chi}^0 $ and $\stop \to b \tilde{\chi}^+$ .

\begin{figure}[p]
\begin{center}
\includegraphics[angle=-90,width=0.65\textwidth]{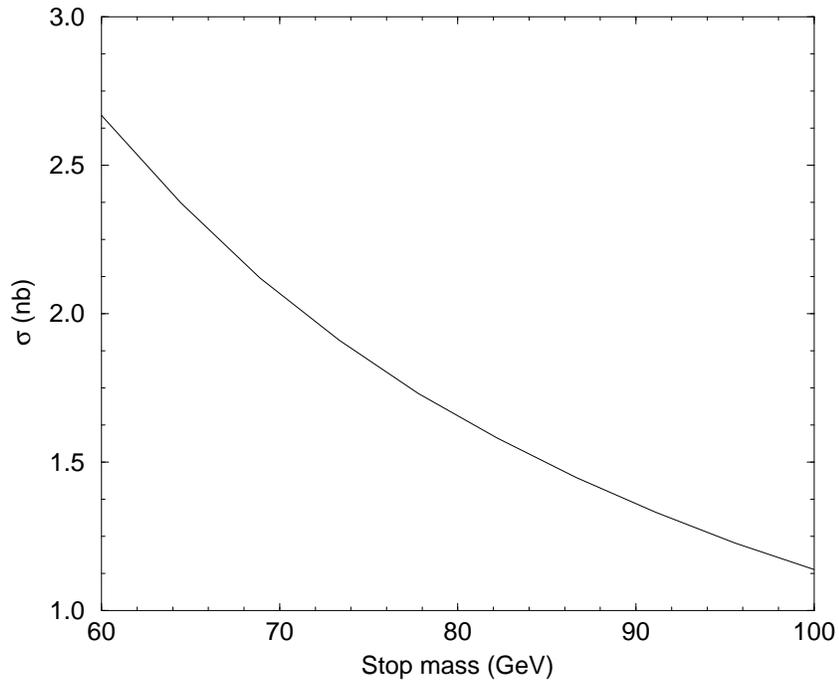}
\end{center}
\caption{ {\em Peak cross section as a function of the stop mass, for the LEP
case, at Born level.}}
\label{fig:stop3}
\end{figure}

\begin{figure}[p]
\begin{center}
\includegraphics[angle=-90,width=0.65\textwidth]{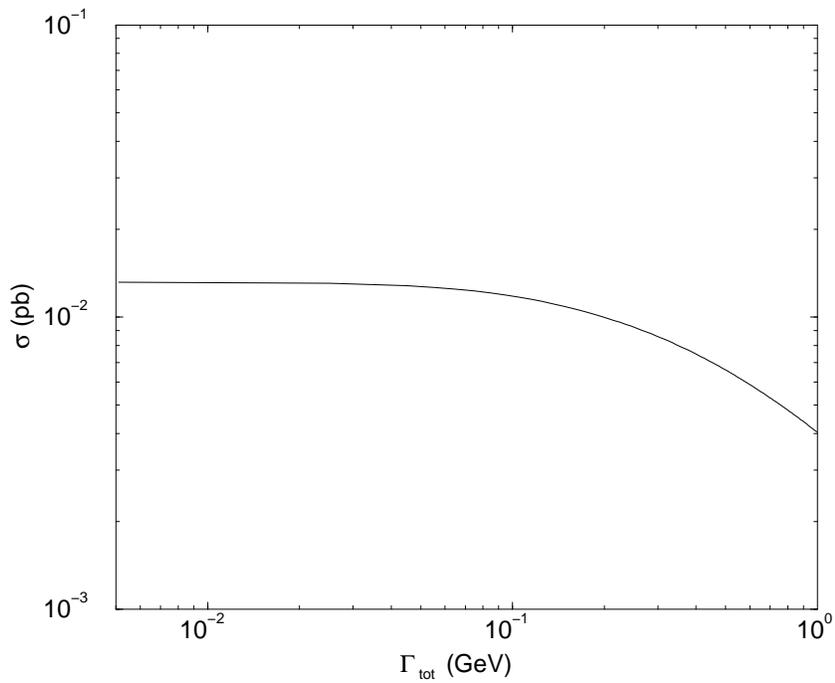}
\end{center}
\caption{ {\em Total cross section folded with a beam energy spread of 200 MeV
as a function of the total width of the stop, at Born level. The plot has 
been obtained for a stop mass of 100 GeV.}}
\label{fig:stop4}
\end{figure}

 Figure \ref{fig:stop3}  shows the peak cross
section obtained from (\ref{bw}) as function of the stop mass for 200 $GeV$
center of mass energy (LEP2). The evaluation of the peak cross--section assumes
that the annihilation modes are dominant.
 While the peak cross section is 
in the $nb$ range, the resonance is practically undetectable at 
the present collider because its width is much smaller than the typical beam
energy spread (of the order of 200 $MeV$ at LEP2 ~\cite{PDG}).   
The effect of a growth of the total width  -- due to e.g. the opening of other
decay channels  \cite{BARTL} -- does not change the result, as 
the net effect will be a decrease of the peak cross section.
This is clearly illustrated in figure \ref{fig:stop4} where the Breit--Wigner
formula (\ref{bw}) is folded with the typical energy spread of the beam of
 200 $MeV$.

\begin{figure}[hb]
\begin{center}
\includegraphics[angle=0,width=0.60\textwidth]{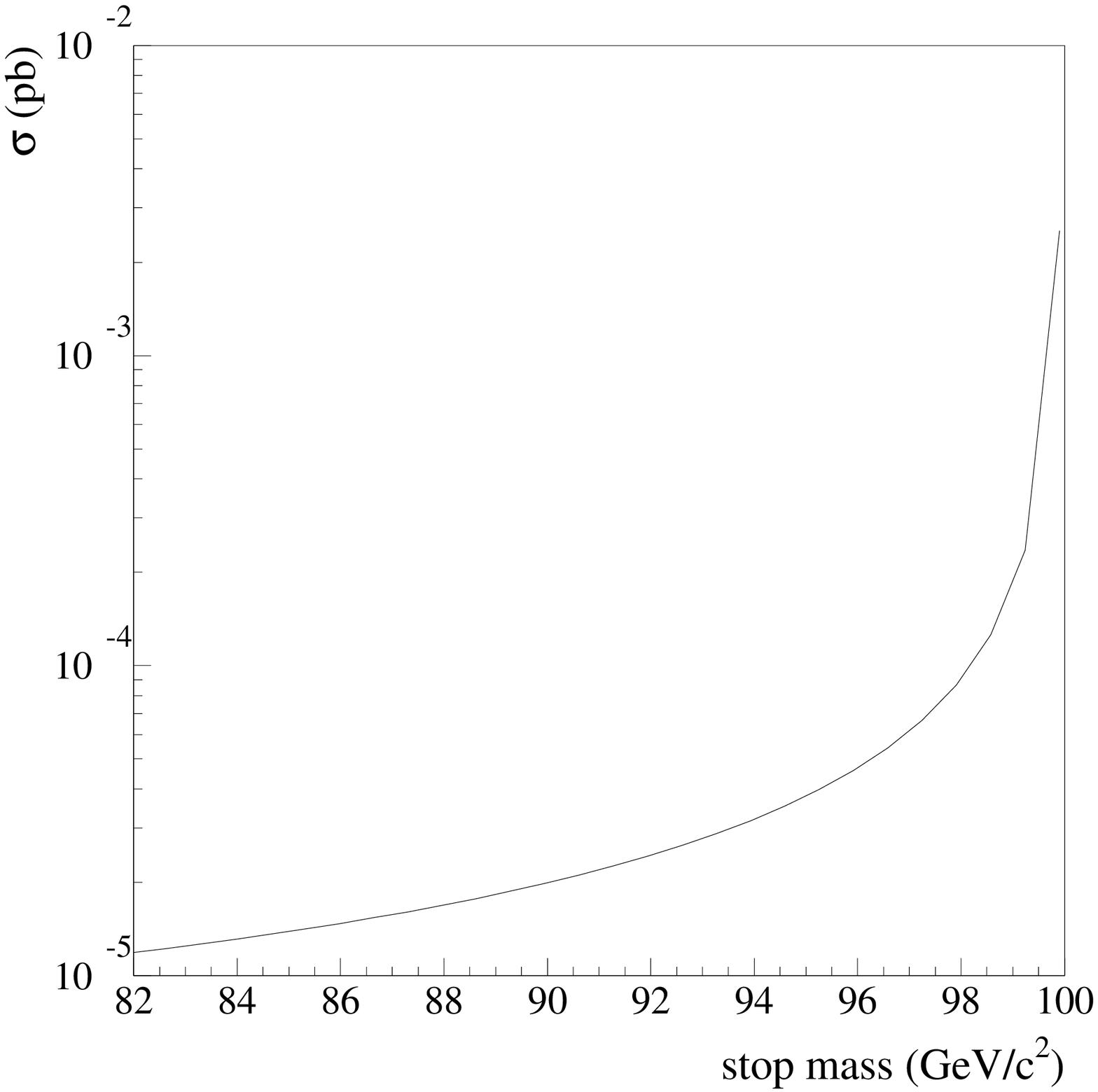}
\end{center}
\caption{ {\em Radiative return production cross section as a function of the 
stop mass, for the LEP case.}}
\label{fig:radia}
\end{figure}

The possibility of stoponium production with radiative returns has also 
been considered and the results for the cross--sections are
illustrated in figure~(\ref{fig:radia}). 
We see that the cross--section is quite small, and in this manner there
is no possibility of seeing any signal.

\begin{figure}[p]
\begin{center}
\includegraphics[angle=-90,width=0.65\textwidth]{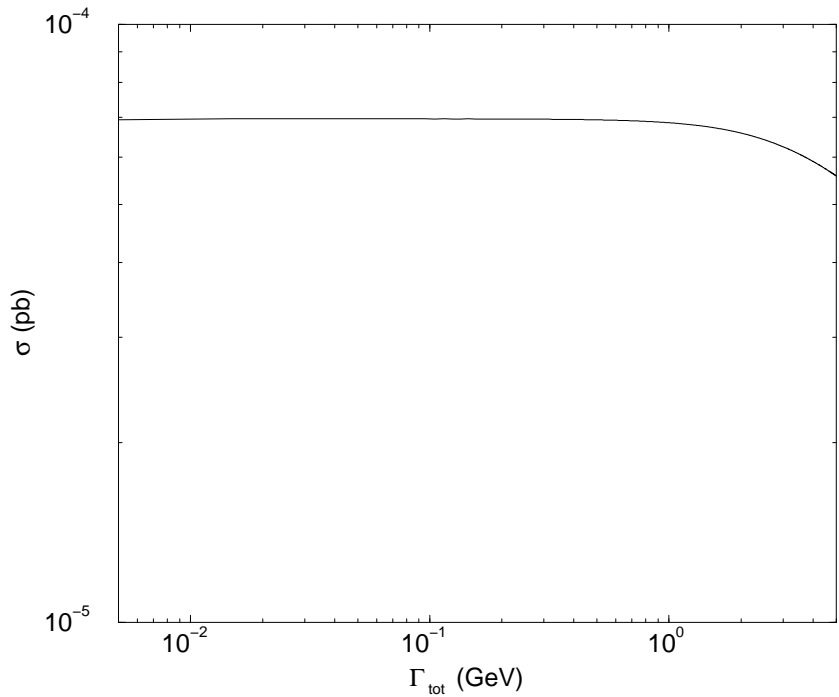}
\end{center}
\caption{ {\em Like Fig.~\ref{fig:stop4}, for a beam energy spread of 
6 GeV (NLC)~\cite{TESLA}.
The plot has been obtained for a stop mass of 200 GeV.}}
\label{fig:stop4.nlc}

\end{figure}

\begin{figure}[hb]
\begin{center}
\includegraphics[angle=0,width=0.65\textwidth]{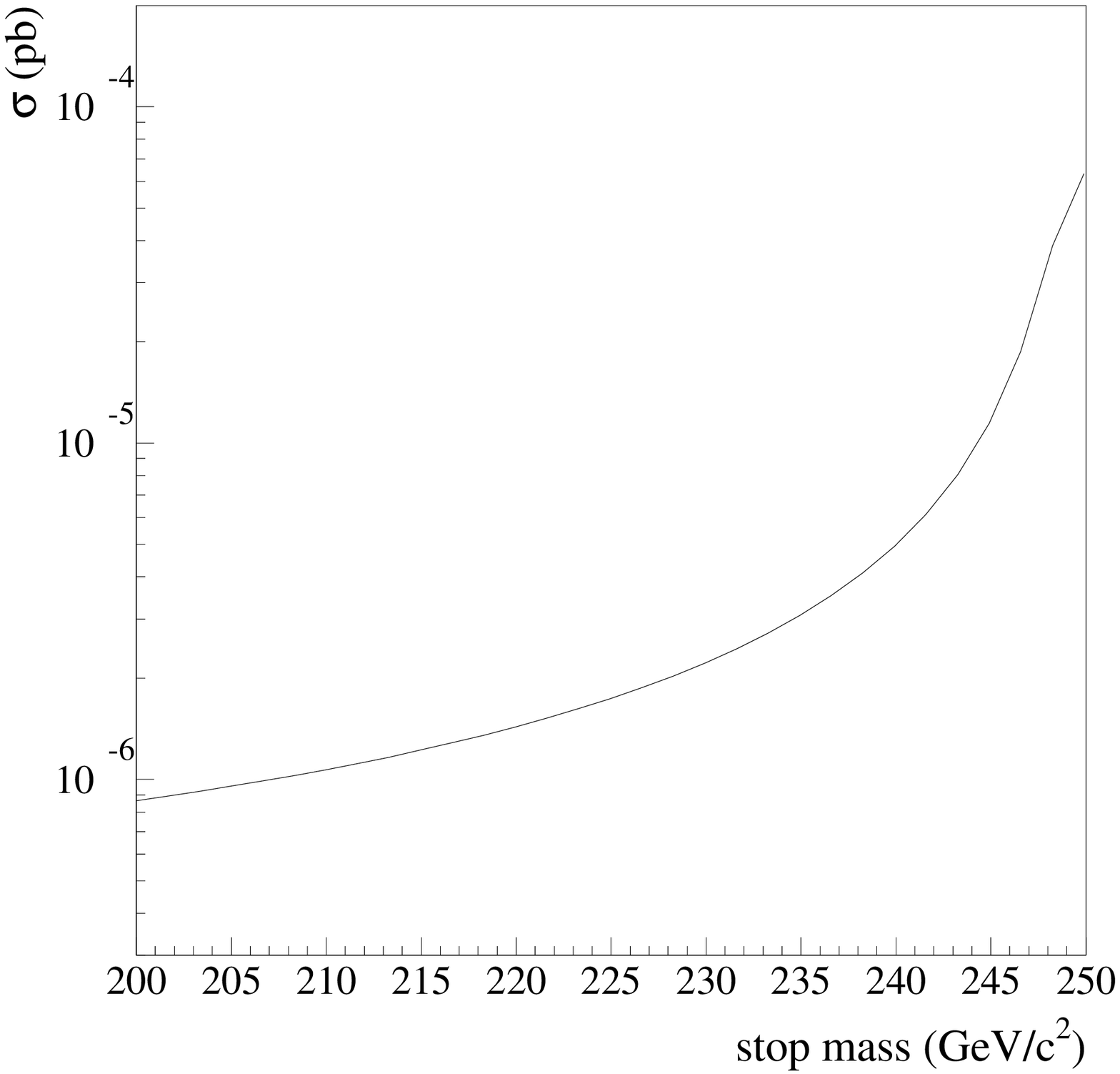}
\end{center}
\caption{ {\em Radiative return production cross section as a function of the
stop mass, for the NLC case.}}
\label{fig:radia.nlc}
\end{figure}

With the increase of the  centre of mass energy (NLC case) the scenario changes:
as more decay channels appear there are regions in the parameter space where 
the stoponium could not be formed. The net result for the signal detection does
not change, as it could clearly be seen in figures~(\ref{fig:stop4.nlc}) 
and~(\ref{fig:radia.nlc}) where we show the effective total cross--section 
and the radiative return cross--section for centre of mass energy of 500 $GeV$.

\section{Conclusions}
We have shown that because of the high energy binding and of the narrow decay
width the formation of a $\stoponium$ $P$ wave bound state is possible 
in certain regions of the parameter space, and in particular for a light stop. 
However our result shows that this supersymmetric bound state cannot be 
detected at the present and even future $e^+e^-$ collider.
The latter fact proves also that it gives a negligible contribution to the 
$\stoponium$ production cross section. \\ \\
{\bf Acknowledgments} \\ 
We Thank G.~Pancheri for useful discussion and suggestions. We would like to
thank also G.~Altarelli and V.~Khoze for careful reading of the manuscript.
One of us (N.F.) wishes to thank A.~Masiero for useful discussion.






\end{document}